\documentclass[twocolumn,amsmath,amssymb,aps]{aastex62}
\usepackage[fleqn]{amsmath}
\usepackage{mathrsfs}

\usepackage{ctable}
\usepackage{graphicx}
\usepackage{dcolumn}
\usepackage{bm}
\def\Rearth{R_\oplus}

\received{February 25, 2020}
\revised{August 7, 2020}
\accepted{August 12, 2020}
\submitjournal{The Astrophysical Journal}

\shorttitle{Rocklines as Cradles for Refractory Solids in the Protosolar Nebula}
\shortauthors{Aguichine et al.}

\begin{document}
	
	\title{Rocklines as Cradles for Refractory Solids in the Protosolar Nebula}

	\author{Artyom Aguichine, Olivier Mousis}
	\email{artem.aguichine@lam.fr}
	\affil{Aix Marseille Univ, CNRS, CNES, LAM, Marseille, France}
	
	\author{Bertrand Devouard}
	\affiliation{Aix Marseille Univ, CNRS, IRD, INRA, Coll France, CEREGE, Aix-en-Provence, France}
	
	\author{Thomas Ronnet}
	\affiliation{Lund Observatory, Department of Astronomy and Theoretical Physics, Lund University, Box 43, 221 00 Lund, Sweden}
	
	\begin{abstract}
		
		In our solar system, terrestrial planets and meteoritical matter exhibit various bulk compositions. To understand this variety of compositions, formation mechanisms of meteorites are usually investigated via a thermodynamic approach that neglect the processes of transport throughout the protosolar nebula. Here, we investigate the role played by rocklines (condensation/sublimation lines of refractory materials) in the innermost regions of the protosolar nebula to compute the composition of particles migrating inward the disk as a function of time. To do so, we utilize a one-dimensional  accretion disk model with a prescription for dust and vapor transport, sublimation and recondensation of refractory materials (ferrosilite, enstatite, fayalite, forsterite, iron sulfide, metal iron and nickel). We find that the diversity of the bulk composition of cosmic spherules, chondrules and chondrites can be explained by their formation close to rocklines, suggesting that solid matter is concentrated in the vicinity of these sublimation/condensation fronts. Although our model relies a lot on the number of considered species and the availability of thermodynamic data governing state changes, it suggests that rocklines played a major role in the formation of small and large bodies in the innermost regions of the protosolar nebula. Our model gives insights on the mechanisms that might have contributed to the formation of Mercury's large core.
		
	\end{abstract}
	
	
	\section{Introduction}
	
	Meteorites and terrestrial planets show varying proportions of silicates and metallic iron, with Fe being distributed between Fe alloys and silicates \citep{Ni00,Fo15}. Chondrites are the most common meteoritical bodies found on Earth, whose unaltered structure gives valuable information on the formation conditions of the building blocks of the solar system \citep{Gr72,Wa88}. 
	Chondrules, which are round grains composed primarily of the silicate minerals olivine and pyroxene and found in the three families of chondrites (ordinary, carbonaceous, and enstatite chondrites), are then important to assess the thermodynamic evolution and the initial composition of the protosolar nebula (PSN) \citep{Co12,Bo17}. Another source of meteoritical matter are cosmic spherules (CS), which are micrometeorites formed by the melting of interplanetary dust particles during atmospheric entry. CS are believed to sample a broader range of material than the collections of meteorites \citep{Ta00}. They are sorted by families and types, each one presenting its own structure and composition \citep{So15,Ta00,Al02,Co11}. Neither chondrules, chondrites, or CS represent the earliest condensates of refractory phases in the PSN, most of which have been thermally processed before accretion on asteroids and planetary bodies. Their composition, however, can reasonably be used as a proxy of the composition of refractory solids in the PSN. The observed variations of composition of chondrites, chondrules, and CS suggest a compositional and redox gradient of refractory matter as a function of radial distance in the PSN and/or time. Recent studies explored the link between CS and chondrites, showing that it is possible to associate CS to their chondritic precursors, based on compositional analysis \citep{Ru15,Gi17}, potentially providing additional constraints to the thermodynamic conditions of the PSN.
	
	
	The question of the origin of the composition differences among chondrites and chondrules is often answered by invoking the cooling of the inner disk where high temperature materials condense first \citep{Gr72,TS01,TS04}. However, the influence of the temperature gradient within an evolving PSN, where more refractory materials form at closer distances from the Sun, has never been investigated to address this issue. Abundances of materials both in solid and gaseous phases are ruled by the chemical status of the disk and by the location of their condensation/sublimation fronts \citep{Dr17,Ob19,Pe19}. Significant increases of the abundances of solid materials can be generated at the location of these transition lines, due to the dynamics of vapors and grains \citep{Cy98,Cy99,Bo10,Al14,De17,Mo19}. 
	These processes potentially explain the metallicity of Jupiter through its formation near the water snowline \citep{Mo19}, as well as the high density of Mercury whose building blocks could have formed in regions where the abundances of Fe-bearing species are prominent \citep{TS04,Va19}.
	
	Here, we use a time-dependent coupled disk/transport model to investigate the role held by rocklines (the concept of snowlines extended to more refractory solids) of the most abundant solids into the shaping of Mg, Fe, and Si abundances profiles in the inner part of the PSN. The radial transport of solid grains through the different rocklines, coupled to the diffusion of vapors, leads to local enrichments or depletions in minerals and imply variations of the Mg-Fe-Si composition of dust grains in the inner regions of the PSN. We discuss our results in light of the relative abundances and compositions of minerals observed in meteoritic matter and planetary bulk compositions. Our approach can be used to derive the formation conditions of the primitive matter in the PSN, and to give insights on the origin of Mercury as well as Super-Mercuries \citep{Sa18,Br19}.
	
	Section \ref{sec:model} summarizes the disk/transport model used to compute the composition and the thermodynamic properties of the PSN. Our results are presented in Section \ref{sec:results}. Section \ref{sec:ccls} is devoted to discussion and conclusions.
	
	\section{Model} \label{sec:model}
	\subsection{Disk evolution} \label{sec:disk-model}
	Our time-dependent PSN model is ruled by the following second-order differential equation \citep{Ly74}: 
	\begin{eqnarray}
	\frac{\partial \Sigma_{\mathrm{g}}}{\partial t} = \frac{3}{r} \frac{\partial}{\partial r} \left[ r^{1/2} \frac{\partial}{\partial r} \left( r^{1/2} \Sigma_{\mathrm{g}} \nu \right)\right]. \label{eqofmotion}
	\end{eqnarray}
	
	\noindent This equation describes the evolution of a viscous accretion disk of surface density $\Sigma_{\mathrm{g}}$ of dynamical viscosity $\nu$, assuming hydrostatic equilibrium in the $z$ direction. This equation can be rewritten as a set of two first-order differential equations coupling the gas surface density $\Sigma_{\mathrm{g}}$ field and mass accretion rate $\dot{M}$:
	\begin{align}
	\frac{\partial \Sigma_{\mathrm{g}}}{\partial t} &= \frac{1}{2\pi r} \frac{\partial \dot{M}}{\partial r}, \label{eqofmotion1}\\
	\dot{M} &= 3\pi \Sigma_{\mathrm{g}}\nu \left(1+2Q\right), \label{eqofmotion2}
	\end{align}
	where $Q=\mathrm{d} \ln (\Sigma_{\mathrm{g}}\nu)/\mathrm{d} \ln(r)$. The first equation is a mass conservation law, and the second one is a diffusion equation. The mass accretion rate can be expressed in terms of the gaz velocity field $v_g$ as $\dot{M}=-2\pi r \Sigma_{\mathrm{g}} v_{\mathrm{g}}$.
	
	The viscosity $\nu$ is computed using the prescription of \cite{Sh73} for $\alpha$-turbulent disks:
	\begin{eqnarray}
	\nu = \alpha \frac{c_{\mathrm{s}}^2}{\Omega_{\mathrm{K}}}, \label{viscosity}
	\end{eqnarray}
	where $\Omega_{\mathrm{K}}=\sqrt{GM_{\odot}/r^3}$ is the keplerian frequency with $G$ the gravitational constant, and $c_{\mathrm{s}}$ is the isothermal sound speed given by
	\begin{eqnarray}
	c_{\mathrm{s}} = \sqrt{\frac{RT}{\mu_{\mathrm{g}}}}. \label{soundspeed}
	\end{eqnarray}
	
	\noindent In this expression, $R$ is the ideal gas constant and $\mu_{\mathrm{g}} = 2.31$ g.mol$^{-1}$ is the gas mean molar mass \citep{Lo09}. In Eq. \ref{viscosity}, $\alpha$ is a non-dimensional parameter measuring the turbulence strength, which also determines the efficiency of viscous heating, hence the temperature of the disk. The value of $\alpha$ typically lies in the range 10$^{-4}$--10$^{-2}$, from models calibrated on disk observations \citep{Ha98,Hu05,De17}.
	
	The midplane temperature $T$ of the disk is computed via the addition of all heating sources, giving the expression \citep{Hu05}:
	
	\begin{eqnarray}
	T^4 = &&\frac{1}{2\sigma_{\mathrm{sb}} } \left(\frac{3}{8} \tau_{\mathrm{R}}+\frac{1}{2 \tau_{\mathrm{P}}}\right) \Sigma_{\mathrm{g}} \nu \Omega_{\mathrm{K}}^2 \nonumber \\
	&+& T_{\odot}^4 \left[ \frac{2}{3\pi} \left(\frac{R_{\odot}}{r}\right)^3 + \frac{1}{2} \left(\frac{R_{\odot}}{r}\right)^2 \left(\frac{H}{r}\right) \left( \frac{\mathrm{d}\ln H }{\mathrm{d}\ln r} - 1\right) \right] \nonumber \\
	&+& T_{\mathrm{amb}}^4~. \label{eq_new_temperature}
	\end{eqnarray}
	
	\noindent The first term corresponds to the viscous heating \citep{Na94}, where $\sigma_{\mathrm{sb}}$ is the Stefan-Boltzmann constant, and $\tau_{\mathrm{R}}$ and $\tau_{\mathrm{P}}$ are the Rosseland and Planck mean optical depths, respectively. For dust grains, we assume $\tau_{\mathrm{P}}=2.4 \tau_{\mathrm{R}}$ \citep{Na94}. $\tau_{\mathrm{R}}$ is derived from \cite{Hu05}:
	
	\begin{eqnarray}
	\tau_{\mathrm{R}} =  \frac{\kappa_{\mathrm{R}} \Sigma_{\mathrm{g}}}{2}, \label{eqrosseland}
	\end{eqnarray}
	
	\noindent where $\kappa_{\mathrm{R}}$ is the Rosseland mean opacity, computed as a sequence of power laws of the form $\kappa_{\mathrm{R}}=\kappa_0 \rho^a T^b$, where parameters $\kappa_0$, $a$ and $b$ are fitted to experimental data in different opacity regimes \citep{Be94} and $\rho$ denotes the gas density at the midplane. The second term corresponds to the irradiation of the disk by the central star of radius $R_\odot$ and surface temperature $T_\odot$. It considers both direct irradiation at the midplane level and irradiation at the surface at a scale height $H=c_{\mathrm{s}}/\Omega_{\mathrm{K}}$. The last term accounts for background radiation of temperature $T_{\mathrm{amb}} = 10$ K.
	
	At each time step, $\Sigma_{\mathrm{g}}$ is evolved with respect to Eq. (\ref{eqofmotion1}). Then Eq. (\ref{eq_new_temperature}) is solved iteratively with Eqs. (\ref{viscosity}), (\ref{soundspeed}), (\ref{eqrosseland}) to produce the new thermodynamic properties of the disk. The new velocity field is then computed following Eq. (\ref{eqofmotion2}). The time-step is computed using the diffusion timescale in each bin $\Delta t = 0.5 \min (\Delta r^2/\nu)\simeq 0.1$ yr, where $\Delta r$ is the spatial grid size, and the factor $0.5$ is taken for safety. The spatial grid is formed from $N=500$ bins scaled as $r_i = (r_\mathrm{min}-r_\mathrm{off})\times i^\beta+r_\mathrm{off}$, where $\beta = \log \left(\frac{r_\mathrm{max}-r_\mathrm{off}}{r_\mathrm{min}-r_\mathrm{off}}\right)/\log(N)$, giving a non-uniform grid, with $r_\mathrm{min}$ and $r_\mathrm{max}$ the limits of the grid, and $r_\mathrm{off}$ is a parameter that allows to control the spatial distribution of points in the simulation box. We compute the sum of the mass lost from both limits of the simulation box ($\int \dot{M}(r_\mathrm{min})dt$ and $\int \dot{M}(r_\mathrm{max})dt$) along with the mass of the disk itself at each time. This quantity remains constant within the precision of the machine.
	
	The initial condition is the self-similar solution $\Sigma_{\mathrm{g}} \nu \propto \exp \left(-(r/r_{\mathrm{c}})^{2-p}\right)$ derived by \cite{Ly74}. Choosing $p=\frac{3}{2}$ for an early disk, we derive the initial surface density and mass accretion rate as follows:
	
	\begin{eqnarray}
	\left\{
	\begin{array}{ll}
	\Sigma_{\mathrm{g},0} = \frac{\dot{M}_{\mathrm{acc},0}}{3\pi \nu} \exp \left[-\left(\frac{r}{r_{\mathrm{c}}}\right)^{0.5}\right], \\
	\dot{M}_0 = \dot{M}_{\mathrm{acc},0} \left(1- \left(\frac{r}{r_{\mathrm{c}}}\right)^{0.5} \right)\exp\left[-\left(\frac{r}{r_{\mathrm{c}}}\right)^{0.5}\right].
	\end{array}
	\right. \label{initial}
	\end{eqnarray}
	
	\noindent To compute the numerical value of $\Sigma_{\mathrm{g},0}$, we solve iteratively equation (\ref{eq_new_temperature}) with the imposed density profile in the first member of equation (\ref{initial}). $r_{\mathrm{c}}$ regulates the size of the disk, and from the second member of equation (\ref{initial}) we see that it matches, at $t=0$ to the centrifugal radius. 
	
\begin{figure}[ht!]
		\resizebox{\hsize}{!}{\includegraphics[angle=0,width=5cm]{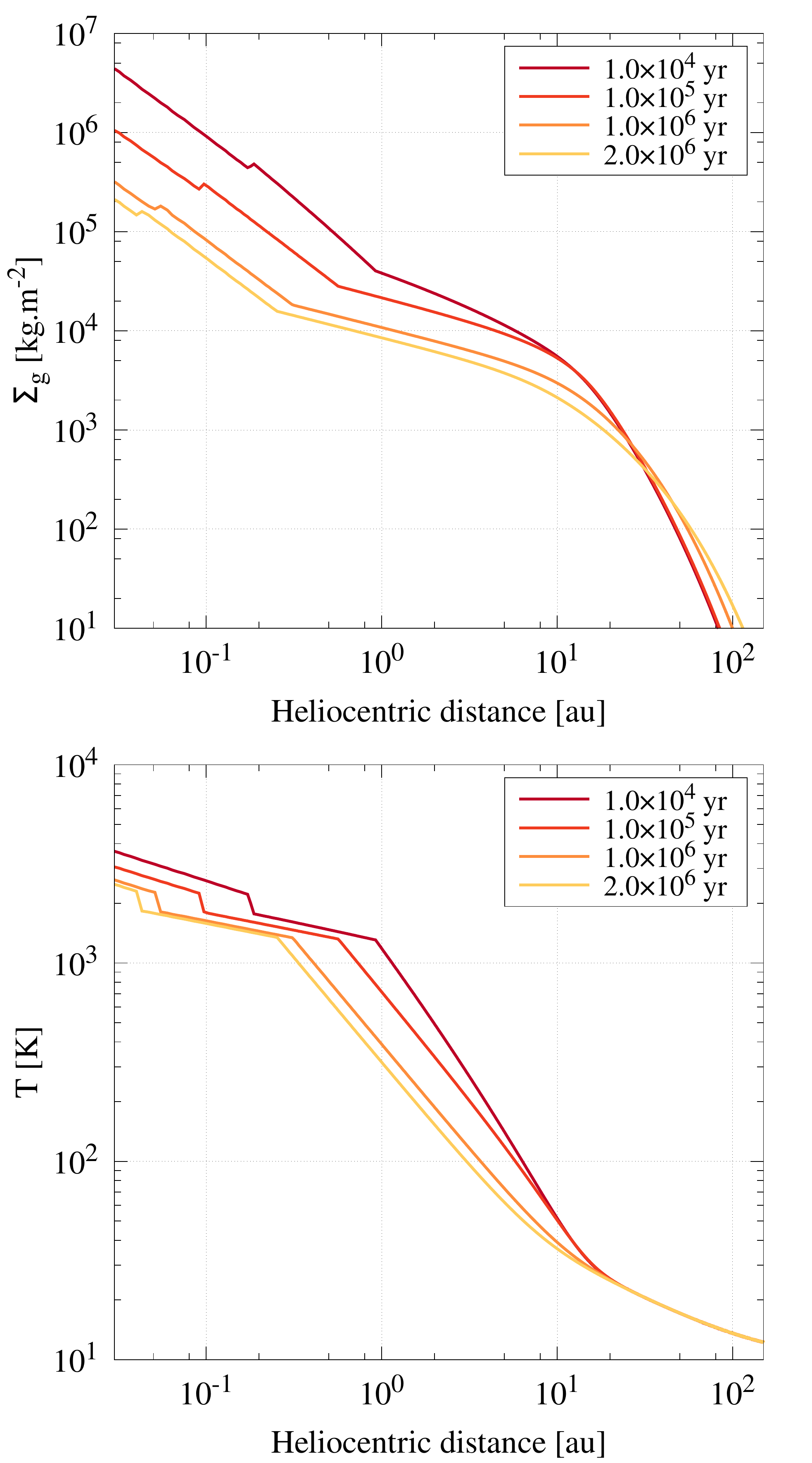}}
		\caption{From top to bottom: disk's surface density and temperature profiles at different epochs of its evolution, assuming $\alpha=10^{-3}$.}
		\label{fig:disk}
	\end{figure}

	The value of $r_{\mathrm{c}}$ is computed by dichotomy, assuming a disk mass $M_d$ of 0.1 M$_\odot$. For $\alpha=10^{-3}$, $r_{\mathrm{c}}$ is equal to 1.83 AU, and 99\% of the disk's mass is encapsulated within $\sim100$ AU. We assume the initial mass accretion rate onto the central star $\dot{M}_{\mathrm{acc},0}$ to be $10^{-7.6}$ M$_\odot \cdot$yr$^{-1}$ \citep{Ha98}. The resulting density and temperature profiles are shown in Figure \ref{fig:disk} at different epochs for $\alpha=10^{-3}$.
	
	\subsection{Size of dust particles}
	
	The size of dust particles used in our model is determined by a two-populations algorithm derived from \cite{Bi12}. This algorithm computes the representative size of particles through the estimate of the limiting Stokes number in various dynamical regimes. We assume that dust is initially present in the form of particles of sizes $a_0 = 10^{-7}$ m, and grow through mutual collisions on a timescale \citep{LJ14}:
	
	\begin{eqnarray}
	\tau_{\mathrm{growth}} = \frac{a}{\dot{a}} = \frac{4\Sigma_{\mathrm{g}}}{\sqrt{3}\epsilon_{\mathrm{g}} \Sigma_{\mathrm{d}} \Omega_{\mathrm{K}}}, \label{taug}
	\end{eqnarray}
	
	\noindent where $a$ is the size of dust grains and $\Sigma_{\mathrm{d}}$ is the dust surface density. We set the growth efficiency parameter $\epsilon_{\mathrm{g}}$ equal to 0.5 \citep{LJ14}. The size of particles that grow through sticking is thus given by $a_{\mathrm{stick}}=a_0 \exp (t/\tau_{\mathrm{growth}})$. However, this growth is limited by several mechanisms preventing particles from reaching sizes greater than $\sim 1$ cm. The first limit arises from fragmentation, when the relative speed between two grains due to their relative turbulent motion exceeds the velocity threshold $u_{\mathrm{f}}$. This sets a first upper limit for the Stokes number of dust grains, which is \citep{Bi12}:
	
	\begin{eqnarray}
	\mathrm{St}_{\mathrm{frag}}= 0.37 \frac{1}{3 \alpha } \frac{u_{\mathrm{f}}^2}{c_{\mathrm{s}}^2}, \label{afrag}
	\end{eqnarray}
	
	\noindent where we set $u_{\mathrm{f}}=10$ m.s$^{-1}$ \citep{Bi12,Mo19}.
	
	\noindent A second limitation for dust growth is due to the drift velocity of grains, i.e. when grains drift faster than they grow, setting an other limit for the Stokes number \citep{Bi12}:
	
	\begin{eqnarray}
	\mathrm{St}_{\mathrm{drift}} = 0.55 \frac{\Sigma_{\mathrm{d}}}{\Sigma_{\mathrm{g}}} \frac{v_{\mathrm{K}}^2}{c_{\mathrm{s}}^2} \left| \frac{\mathrm{d}\ln P}{\mathrm{d} \ln r}\right|^{-1}, \label{adrift}
	\end{eqnarray}
	
	\noindent where $v_{\mathrm{K}}$ is the keplerian velocity, and $P$ is the disk midplane pressure.
	
	Equation (\ref{afrag}) only considers the relative turbulent motion between grains that are at the same location, but the fragmentation threshold $u_\mathrm{f}$ can also be reached when dust grains drift at great velocities, and in the process collide with dust grains that are on their path. In that case, we obtain a third limitation for grains' size \citep{Bi12}:
	
	\begin{eqnarray}
	\mathrm{St}_{\mathrm{df}} = \frac{1}{(1-N)}\frac{u_{\mathrm{f}} v_{\mathrm{K}}}{c_{\mathrm{s}}^2} \left( \frac{\mathrm{d}\ln P}{\mathrm{d}\ln r}\right)^{-1} \label{adf},
	\end{eqnarray}
	
	\noindent where the factor $N=0.5$ accounts for the fact that only particles of bigger size fragment during collisions.
	
\begin{figure}[ht!]
		\resizebox{\hsize}{!}{\includegraphics[angle=0,width=5cm]{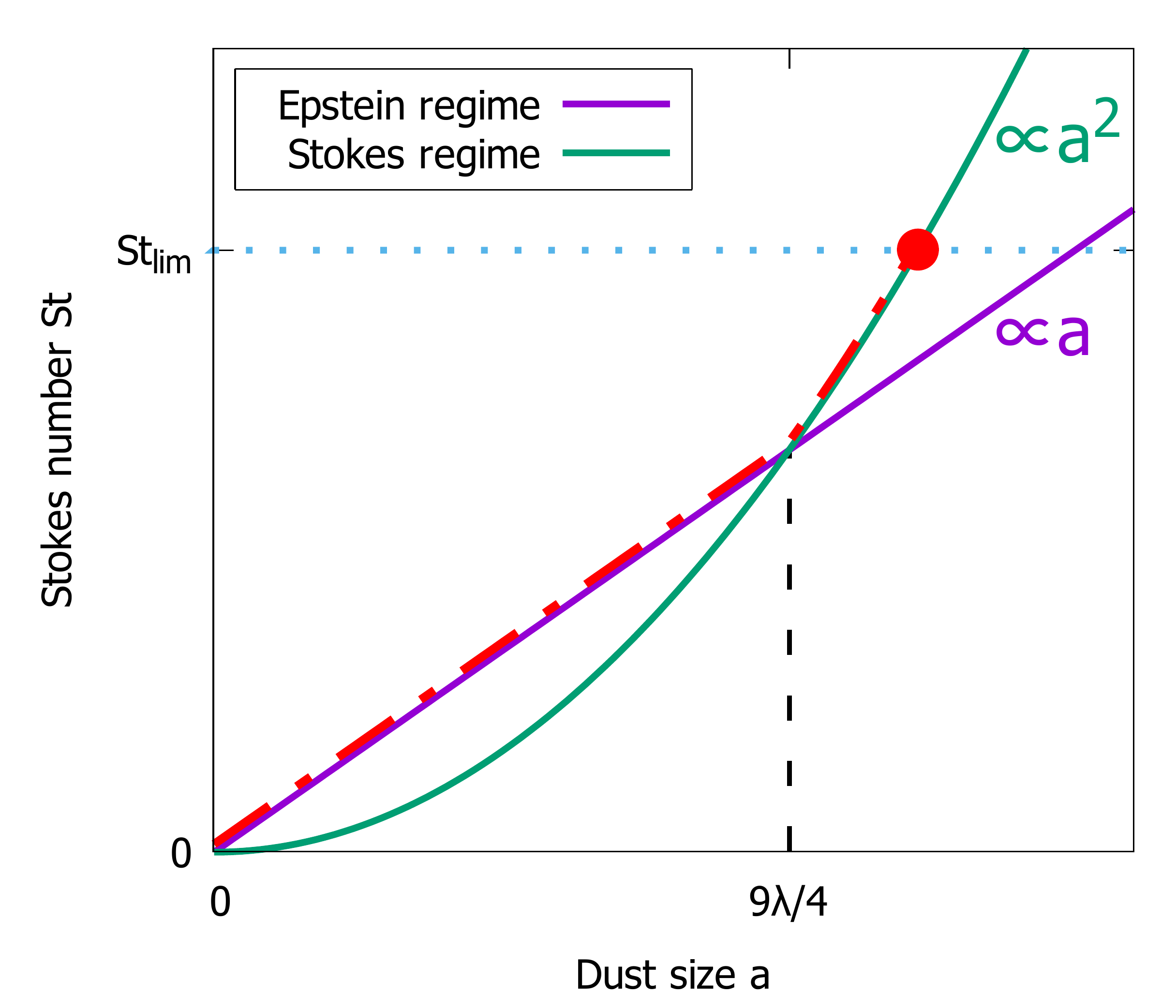}}
		\caption{Visual sketch showing the Stokes number St dependency with respect to the particle size $a$, in both considered flow regimes. When $a<9\lambda/4$, particles follow Epstein regime. When $a>9\lambda/4$, particles follow Stokes regime. In both cases, the relevant size is the smallest among the two, and the relevant Stokes number is the greatest, which are depicted as the red dot-dashed line.}
		\label{fig:st_sketch}
	\end{figure}

	The relation between Stokes number and dust grains size depends on the flow regime in the disk \citep{Jo14}:
	
	\begin{eqnarray}
	\mathrm{St} = \left\{
	\begin{array}{ll}
	\sqrt{2\pi} \frac{a \rho_{\mathrm{b}}}{\Sigma_{\mathrm{g}}} & \text{ if } a \le \frac{9}{4}\lambda\\
	\frac{8}{9} \frac{a^2 \rho_{\mathrm{b}} c_\mathrm{s}}{\Sigma_{\mathrm{g}} \nu} & \text{ if } a \ge \frac{9}{4}\lambda ,
	\end{array}
	\right. \label{St_regimes}
	\end{eqnarray}
	
	\noindent where $\rho_{\mathrm{b}}$ is the bulk density of grains. The first case correspond to the Epstein's regime, occuring in the outermost region of the disk, and the second case corresponds to the Stokes regime. The limit between the two regimes is set by the mean free path $\lambda=\sqrt{\pi/2}~\nu/c_\mathrm{s}$ (computed by equating both terms of Eq. (\ref{St_regimes})) in the midplane of the disk. Combining Eqs. (\ref{taug}-\ref{St_regimes}),  the limiting Stokes number and representative particle size are computed for both regimes independently, giving $a_\mathrm{E}$ and $\mathrm{St}_\mathrm{E}$ in Epstein regime, and $a_\mathrm{S}$ and $\mathrm{St}_\mathrm{S}$ in Stokes regime. Then, as shown by Figure \ref{fig:st_sketch}, the representative size and Stokes number is given by
	
	\begin{align}
	a &= \min \left(a_\mathrm{E},a_\mathrm{S}\right), \label{final_size} \\
	\mathrm{St} &= \max \left(\mathrm{St}_\mathrm{E},\mathrm{St}_\mathrm{S}\right). \label{final_stokes}
	\end{align}
	In the following, two end-cases are considered. In {\it case A}, we assume that all trace species are entirely independent, i.e. a run with several species is equivalent to several runs with a single traces species at a time. In {\it case B}, we assume that at each orbital distance, dust grains are a mixture of all available solid matter at that distance. For this case, the considered dust surface density is the sum over all surface densities $\Sigma_{\mathrm{d}} = \sum_{i} \Sigma_{\mathrm{d},i}$. The bulk density $\rho_\mathrm{b}$ of resulting grains is also the mass-average of the bulk densities of its constituents (given in Table \ref{table_species}):
	\begin{eqnarray}
	\rho_\mathrm{b} = \frac{\sum_{i} \Sigma_{\mathrm{d},i} \rho_{\mathrm{b},i}}{\sum_{i} \Sigma_{\mathrm{d},i}}.
	\end{eqnarray}
	{\it Case B} is favored from a dynamical point of view, as dust grains of different composition mix over long timescales, whereas {\it case A} is favored from a thermodynamic point of view, since sublimation and condensation tend to separate species into their pure forms. Since we focus on rocklines, whose positions are all at distances $\leq 1$ AU, we consider our disk volatile-free. In our model, the closest iceline would be that of H$_2$O, which is located at $\sim$4 AU. This iceline is far enough to ignore its impact on processes at play around rocklines.
	
	\subsection{Evolution of vapors and dust} \label{sec:evol_vapdust}
	
	We follow the approaches of \cite{De17} and \cite{Dr17} for the dynamics of trace species in term of motion and thermodynamics, respectively. The disk is uniformly filled with seven refractory species considered dominant (see Table \ref{table_species}), assuming protosolar abundances for Fe, Mg, Ni, Si and S \citep{Lo09}, similar Fe/Mg ratios in olivine and pyroxene, and that half of Ni is in pure metallic form while the remaining half is in kamacite.
	
\begin{figure}[ht!]
		\resizebox{\hsize}{!}{\includegraphics[angle=0,width=5cm]{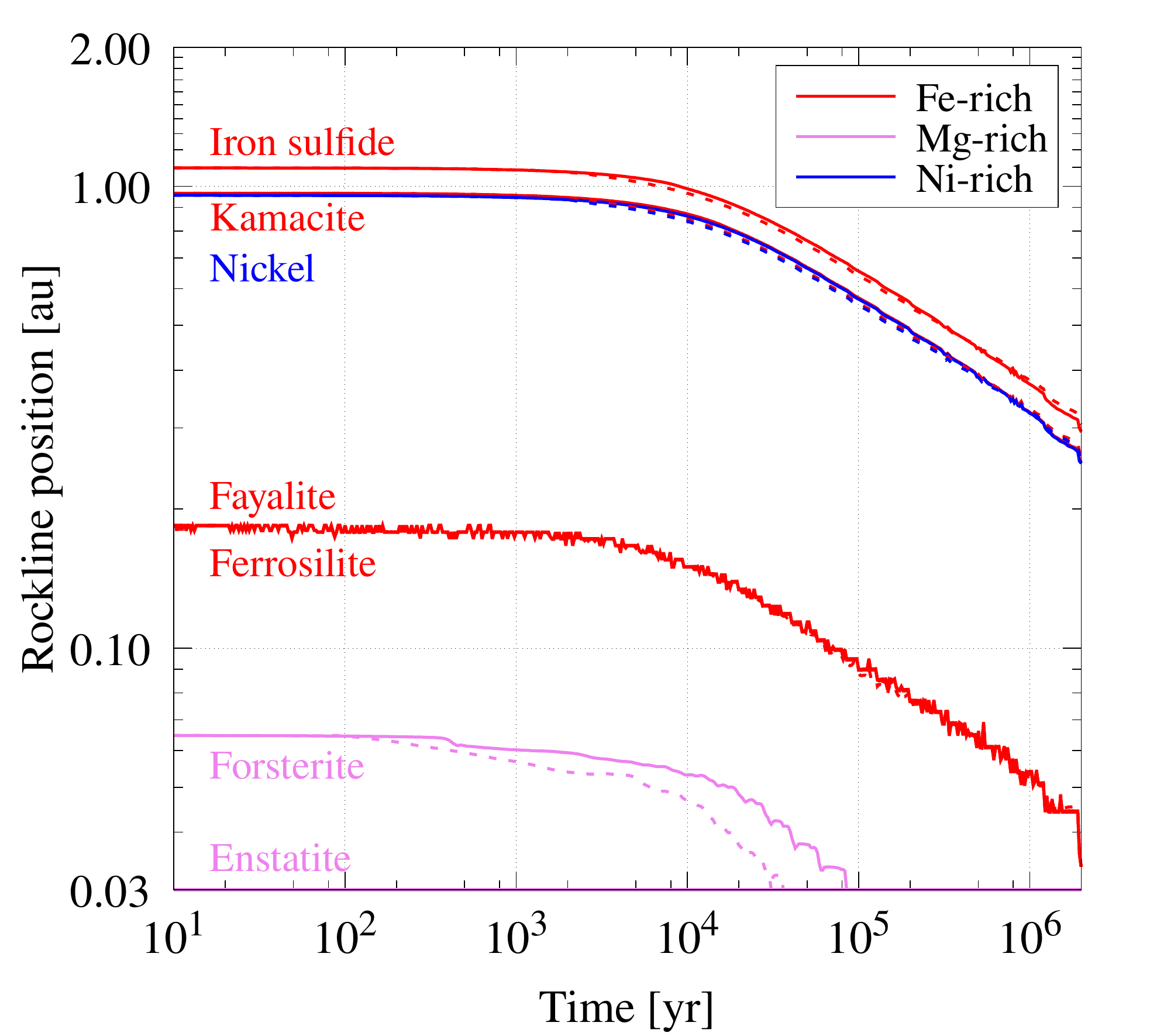}}
		\caption{Time evolution of the locations of rocklines in the PSN. Solid and dashed lines correspond to cases A and B, respectively (see text). Only minor differences between the two cases are observed, resulting from changes in radial drift velocities of particles.}
		\label{fig:rocklines}
	\end{figure}

	Sublimation of grains occurs during their inward drift when partial pressures of trace species become lower than the corresponding vapor pressures. Once released, vapors diffuse both inward and outward. Because of the outward diffusion, vapors can recondense back in solid form, and condensation occurs either until thermodynamic equilibrium is reached or until no more gas is available to condense. Over one integration time step $\Delta t$, the amount of sublimated or condensed matter is \citep{Dr17}
	
	\begin{align}
	\Delta \Sigma_{\mathrm{subl},i} &= \dot{Q}_{\mathrm{subl},i} \Delta t \nonumber \\
	&= \mathrm{min}\left(\frac{6\sqrt{2\pi}}{\pi \rho_{\mathrm{b}} a} \sqrt{\frac{\mu_i}{RT}} P_{\mathrm{sat},i} \Sigma_{\mathrm{d},i}\Delta t~;~\Sigma_{\mathrm{d},i} \right),
	\end{align}
	
	\begin{align}
	\Delta \Sigma_{\mathrm{cond},} &= \dot{Q}_{\mathrm{cond},i} \Delta t \nonumber \\
	&= \min \left( \frac{2 H \mu_g}{R T} \cdot\left(P_{\mathrm{v},i}-P_{\mathrm{sat},i}\right)~;~ \Sigma_{\mathrm{v},i}\right),
	\end{align}
	
	\noindent where $\mu_i$ is the molar mass of a given trace species, $P_{\mathrm{sat},i}$ its saturation pressure and $P_{\mathrm{v},i}$ its partial pressure at a given time and place in the PSN. The second term of the min function ensures that the amount of sublimated (resp. condensed) matter is at most the amount of available matter to sublimate (resp. condense) i.e. the dust (resp. vapor) surface density $\Sigma_{\mathrm{d},i}$ (resp. $\Sigma_{\mathrm{v},i}$). We define the rockline as the location at which the surface density $\Sigma_{\mathrm{d},i}$ of solid and $\Sigma_{\mathrm{v},i}$ and vapor of a given species $i$ are equal. Species exist mostly in solid forms at greater heliocentric distances than their rocklines while they essentially form vapors at distances closer to the central star. Figure \ref{fig:rocklines} shows the locations of the considered rocklines as a function of time in the PSN for both cases. No gas phase chemistry is assumed in the disk.
	
	The motion of dust and vapor, who coexist as separate surface densities $\Sigma_{\mathrm{d},i}$ and $\Sigma_{\mathrm{v},i}$, is computed by integrating the 1D radial advection-diffusion equation \citep{Bi12,De17}:
	
	\begin{equation}
	\frac{\partial \Sigma_{i}}{\partial t}+\frac{1}{r} \frac{\partial}{\partial r}\left[r\left(\Sigma_{i} v_{i}-D_{i} \Sigma_{\mathrm{g}} \frac{\partial}{\partial r}\left(\frac{\Sigma_{i}}{\Sigma_{\mathrm{g}}}\right)\right)\right]+\dot{Q}_{i}=0.
	\end{equation}
	
	\noindent This equation holds for both vapor and solid phases since the motion is determined by the radial velocity $v_i$ and the radial diffusion coefficient $D_i$ of species $i$, $\dot{Q}$ being the source/sink term. When a species $i$ is in vapor form, we assume $v_i \simeq v_g$ and $D_i \simeq \nu$. When this species is in solid form, the dust radial velocity is the sum of the gas drag induced velocity and the drift velocity \citep{Bi12}:
	
	\begin{eqnarray}
	v_{\mathrm{d}} = \frac{1}{1+ \mathrm{St}^2}v_{\mathrm{g}} + \frac{ 2\mathrm{St}}{1+ \mathrm{St}^2}  v_{\mathrm{drift}}, \label{dust_velocity}
	\end{eqnarray}
	
	\noindent where the drift velocity is given by \citep{We77,Na86}:
	
	\begin{eqnarray}
	v_{\mathrm{drift}} = \frac{c_{\mathrm{s}}^2}{v_{\mathrm{K}}} \frac{\mathrm{d}\ln P}{\mathrm{d}\ln r}.
	\end{eqnarray}
	
	\indent This expression usually holds for a population of particles sharing the same size. However, we work here with the two-population algorithm from \cite{Bi12}, i.e. the dust is composed of a mass fraction $f_\mathrm{m}$ of particles of size $a$, and a mass fraction $1-f_\mathrm{m}$ of particles of size $a_0$. The dust radial velocity can be then approximated by a mass-weighted velocity:
	
	\begin{eqnarray}
	v_i = f_\mathrm{m} v_{d,\text{size }a} + (1-f_\mathrm{m}) v_{d,\text{size }a_0},
	\end{eqnarray}
	
	\noindent where $v_{d,\text{size }a/a_0}$ is calculated with respect to Eq. (\ref{dust_velocity}), St is computed for both populations at each heliocentric distance, and $f_{\mathrm{m}}$ depends on the size limiting mechanism \citep{Bi12}:
	
	\begin{eqnarray}
	f_{\mathrm{m}} = \left\{
	\begin{array}{ll}
	0.97 & \text{ if } \mathrm{St}_{\mathrm{drift}} =\text{min}\left(\mathrm{St}_{\mathrm{frag}}, \mathrm{St}_{\mathrm{drift}}, \mathrm{St}_{\mathrm{df}}\right)\\
	0.75 & \text{ otherwise.}
	\end{array}
	\right. \nonumber
	\end{eqnarray}
	
	\noindent Due to the low Stokes number of dust grains ($\mathrm{St}<1$), we make the approximation $D_i = \frac{\nu}{1+\mathrm{St}^2_i} \simeq \nu$.
	
	\begin{table*}
		\centering
		\caption{Main refractory phases present in the disk with corresponding initial abundances and references for saturation pressures $P_{\mathrm{sat}}$.} 
		\label{table_species}
		\begin{tabular}{lllccc}
			\tablewidth{0pt}
			\hline
			\hline
			Type 	& Name 		& Formula 			& n$_\mathrm{i}$ = (X$_\mathrm{i}$/H$_2$)				& $\rho_b$ (g.cm$^{-3}$)	& $P_{\mathrm{sat}}$  \\
			\hline
			Pyroxene &Ferrosilite 	& FeSiO$_3$    		& $9.868\times 10^{-6}$     	&3.95		& \cite{Nag94} \\
			&Enstatite 	&  MgSiO$_3$   		& $4.931\times 10^{-5}$		&3.20		& \cite{Ta02} \\
			Olivine  	&Fayalite 		& Fe$_2$SiO$_4$  		& $2.897\times 10^{-6}$  		&4.39		& \cite{Nag94} \\
			&Forsterite 	& Mg$_2$SiO$_4$ 		&  $1.452\times 10^{-5}$ 		&3.22		& \cite{Nag94} \\
			Alloys   	&Iron sulfide 	&  FeS 				& $3.266\times 10^{-5}$ 		&4.75		& \cite{Fe89} \\
			&Kamacite 	&  Fe$_{0.9}$Ni$_{0.1}$   &   $1.875\times 10^{-5}$     	&7.90		& \cite{Al84} \\
			&Nickel 		&  Ni    				& $1.875\times 10^{-6}$ 		&8.90		& \cite{Al84} \\
			\hline
		\end{tabular}
	\end{table*}
	
	\section{Results} \label{sec:results}
	
\begin{figure*}[ht!]
		\resizebox{\hsize}{!}{\includegraphics[angle=0,width=5cm,trim=0.7cm 1.2cm 0.9cm 1cm,clip]{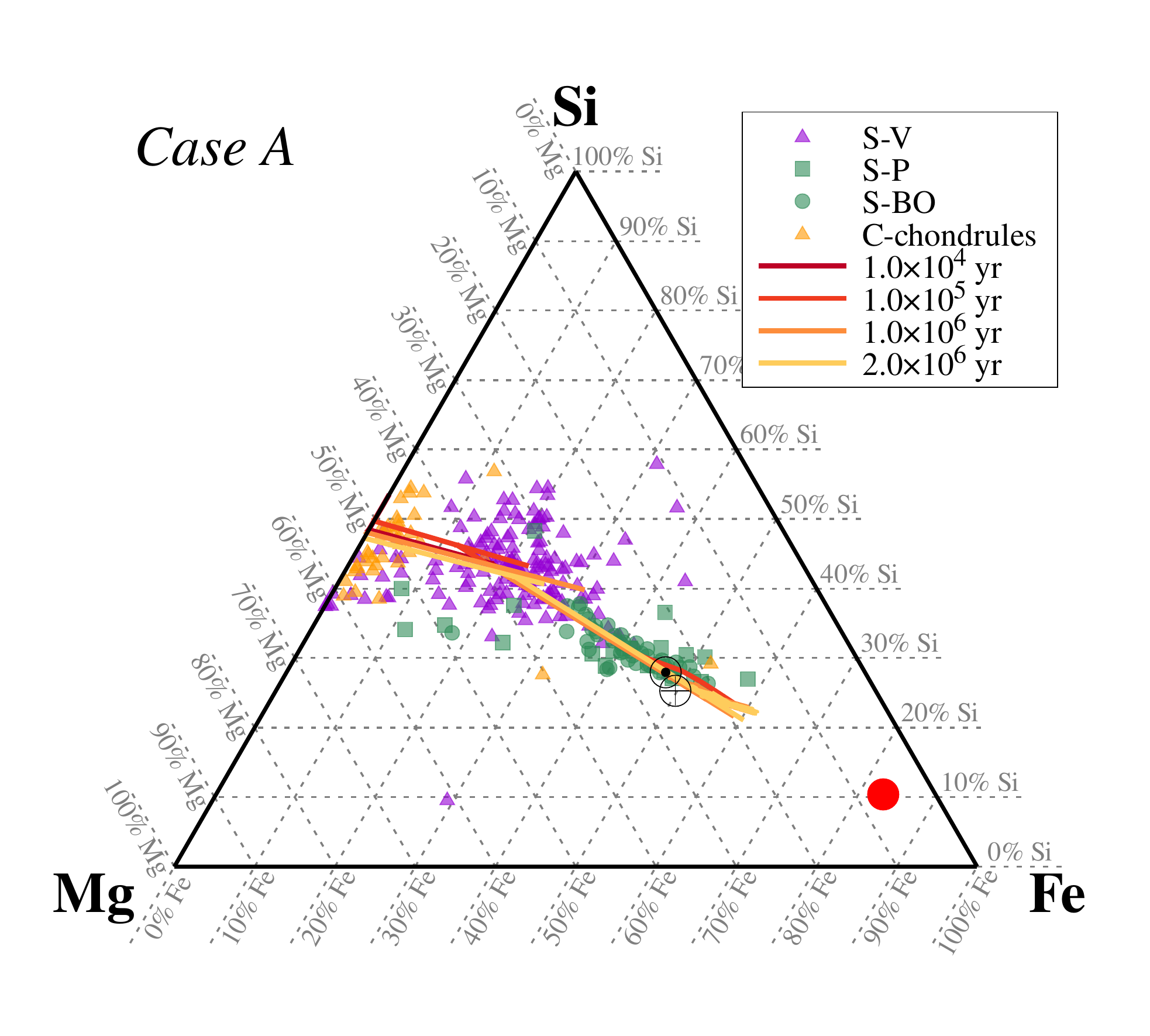}\includegraphics[angle=0,width=5cm,trim=0.0cm 0cm 0cm 0cm,clip]{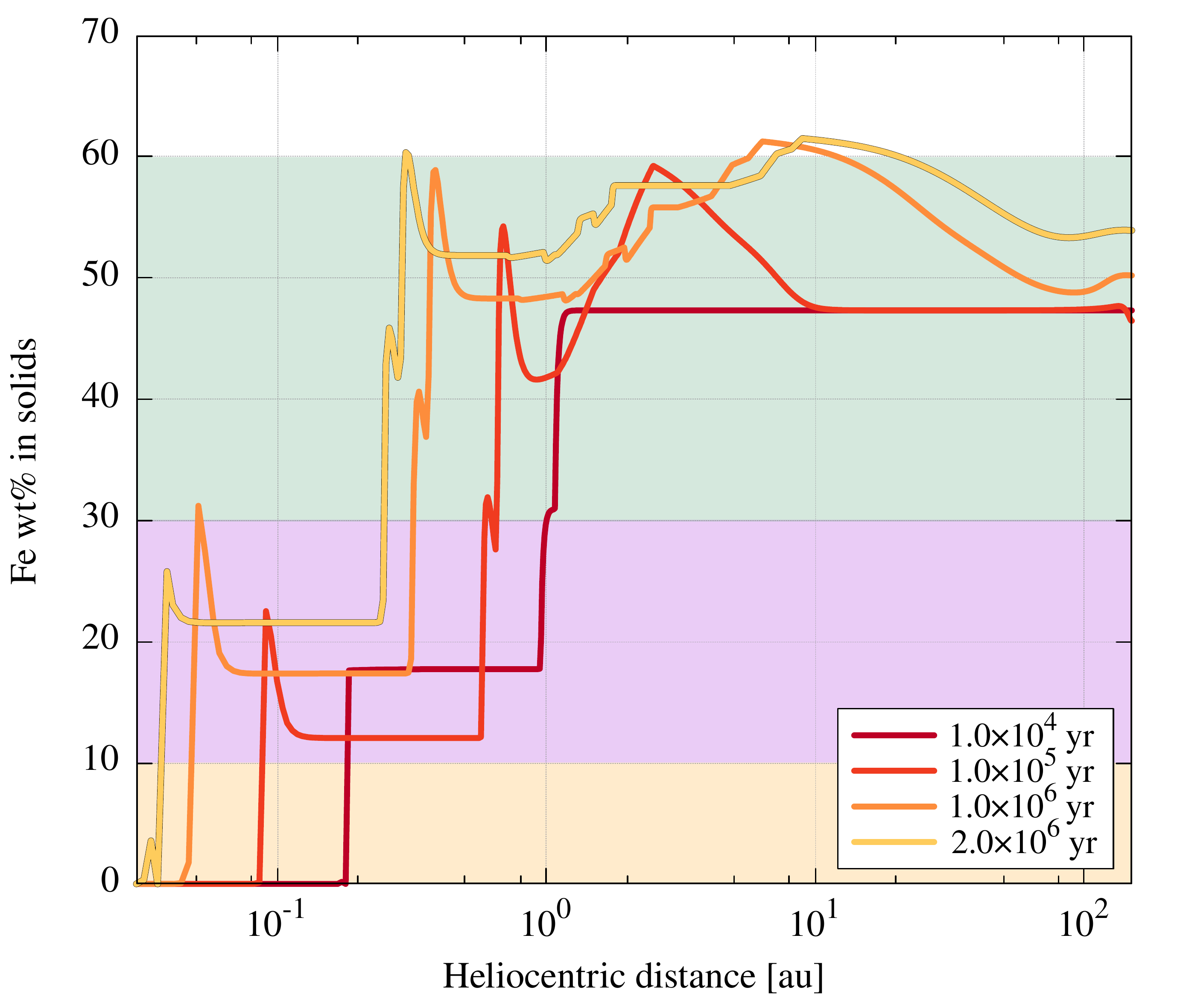}}
		\caption{Left panel: composition of refractory matter in a Mg-Fe-Si ternary diagram expressed in mass fraction between $t~=~10^4$ and $2~\times~10^6$ yr of the PSN evolution with $\alpha=10^{-3}$ and for \textit{case A} (all trace species independent). Purple triangles correspond to glass cosmic spherules (S-V type) \citep{Ta00}, suggesting they were formed by condensation in the vicinities of Fe oxides rocklines. Green circles correspond to barred olivine spherules (S-BO type) \citep{Co11} that potentially formed via mixing in the region of iron alloys rocklines. Green squares represent porphyritic spherules (S-P type) from the same collection. Yellow triangles correspond to a random selection of chondrules from various carbonaceous chondrites studied in \cite{He10}. Sun and Earth symbols correspond to protosolar and Earth bulk compositions, respectively \citep{So07}. The red circle represents Mercury's bulk composition \citep{Br18}. Right panel: evolution of the iron abundance (in wt\%) in solids as a function of heliocentric distance. The different colorboxes correspond to the iron content in chondrules (0--10\%), glass cosmic spherules (10--30\%), and porphyritic and barred olivine cosmic spherules (30--60\%).}
		\label{fig:ternaryA}
	\end{figure*}

\begin{figure*}[ht!]
		\resizebox{\hsize}{!}{\includegraphics[angle=0,width=5cm,trim=0.7cm 1.2cm 0.9cm 1cm,clip]{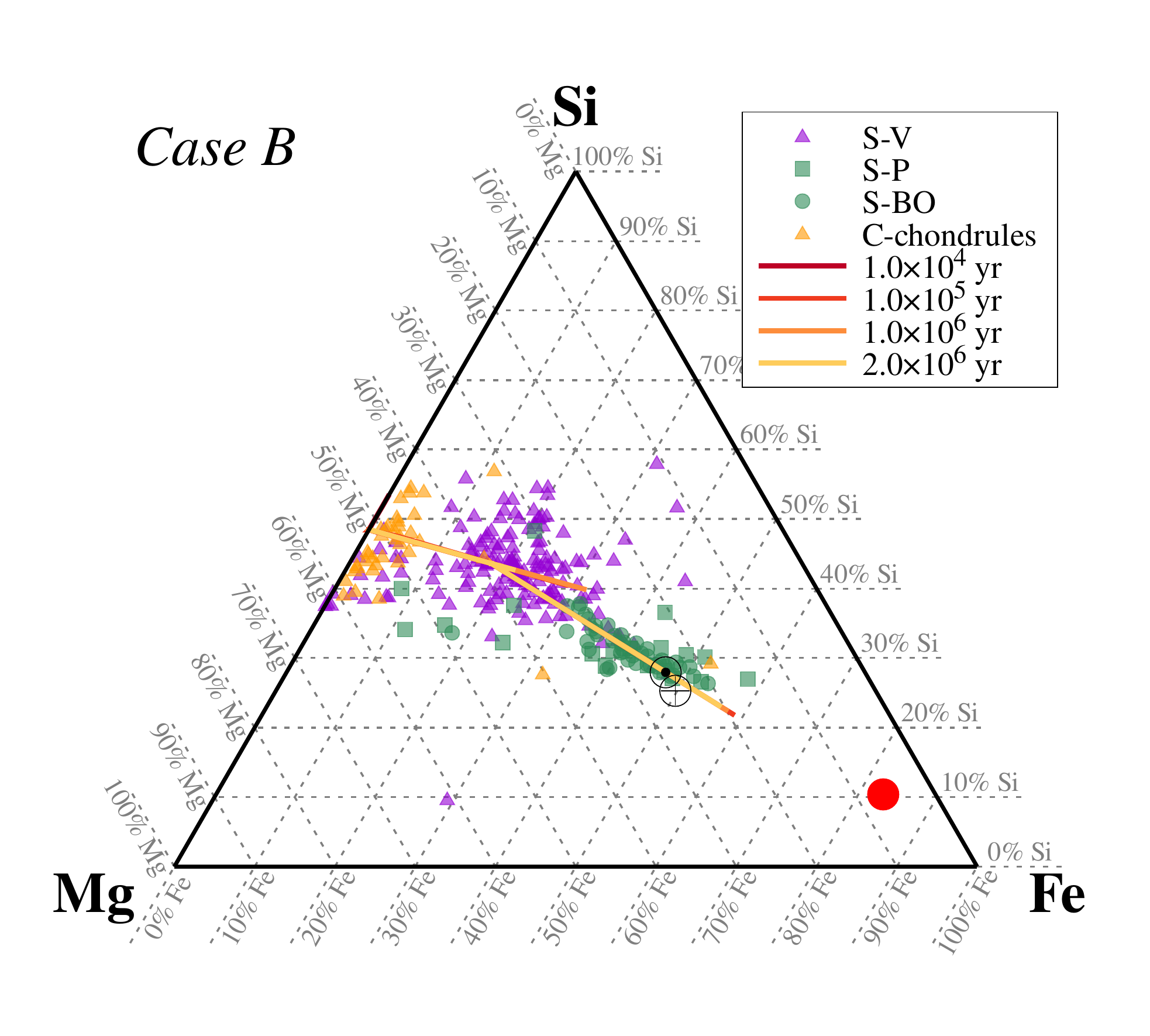}\includegraphics[angle=0,width=5cm,clip]{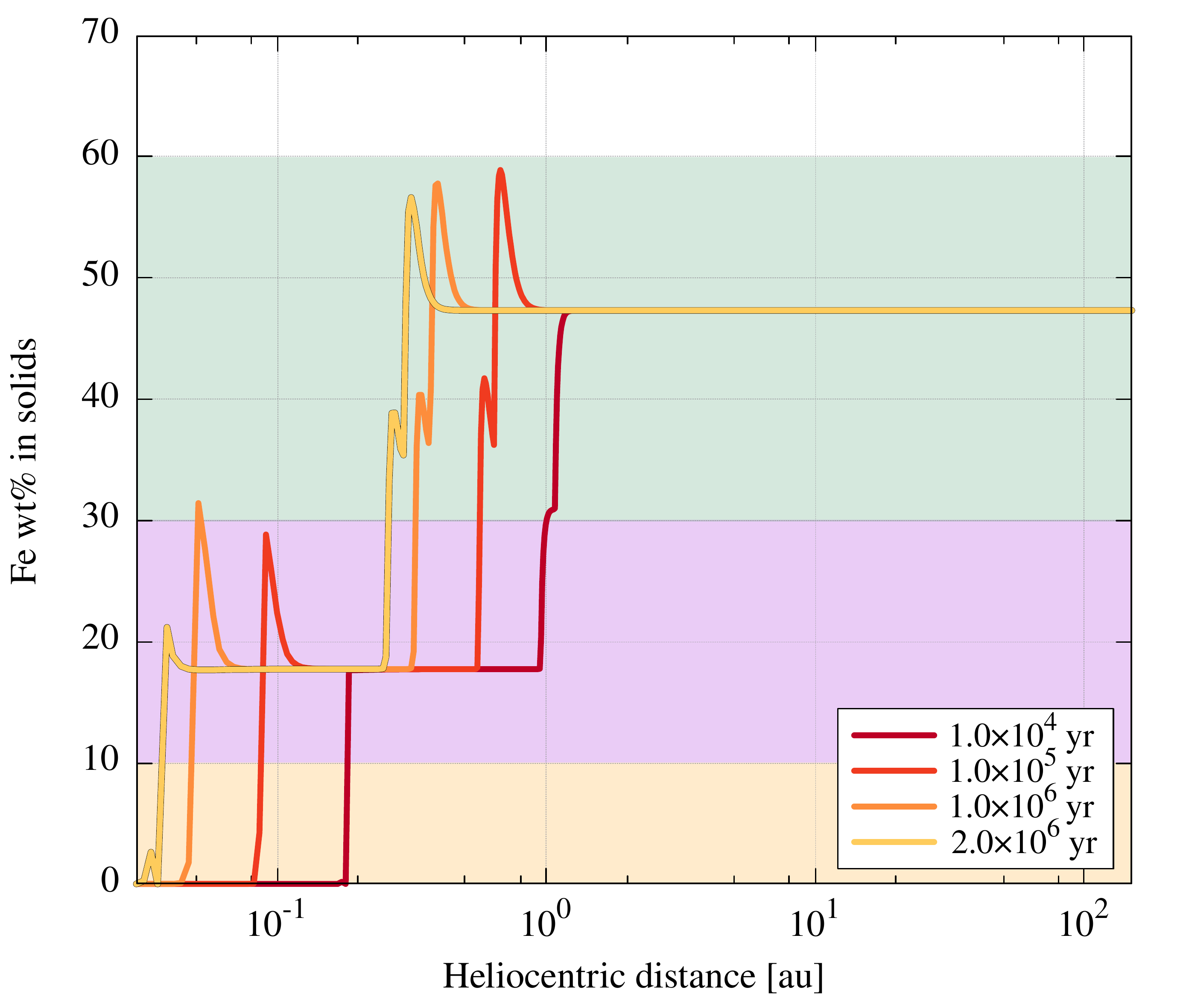}}
		\caption{Same as figure \ref{fig:ternaryA}, but for \textit{case B}.}
		\label{fig:ternaryB}
	\end{figure*}

	Figures \ref{fig:ternaryA} and \ref{fig:ternaryB} represent the time evolution of the composition of refractory particles evolving throughout the PSN in a Mg-Fe-Si ternary diagram and the time evolution of the Fe abundance profile (in wt\%) as a function of heliocentric distance, and in {\it case A} and {\it case B}, respectively. Ternary diagrams display the composition of cosmic spherules and chondrules. Solid particles start with a protosolar composition which changes during their drift throughout the innermost regions of the PSN, due to the successive sublimation of minerals. Because alloys, which contain only Fe, S and Ni, are the first to sublimate (see Fig. \ref{fig:rocklines}), solid particles loose a substantial amount of iron (in the $\sim$20--50$\%$ range), but not Mg and Si. This is illustrated by the composition profiles that shift toward the Mg-Si axis with an unchanged Mg/Si ratio on the ternary diagrams. Closer to the Sun, ferrosilite and fayalite begin to sublimate until no Fe remains in solid form. In this case, the composition profiles in ternary diagrams are located on the Mg-Si axis, and because some silicon is vaporized too, the Mg/Si ratio increases. Finally, if the temperature is high enough to sublimate forsterite, only enstatite remains in solid particles, with an atomic ratio $(\mathrm{Mg/Si})_{\mathrm{at}} = 1$ corresponding to 46 wt\% of Mg. In {\it case A}, lighter grains are more subject to the pressure gradient and drift inward faster than dense Fe-rich grains. As a result, the Fe wt\% increases by $\sim 10\%$ in the whole disk. This leads to a slightly wider range of possible compositions in the PSN than for {\it case B}, but the differences between the two runs are minor from a compositional point of view.
	
\begin{figure}[ht!]
		\resizebox{\hsize}{!}{\includegraphics[angle=0,width=5cm]{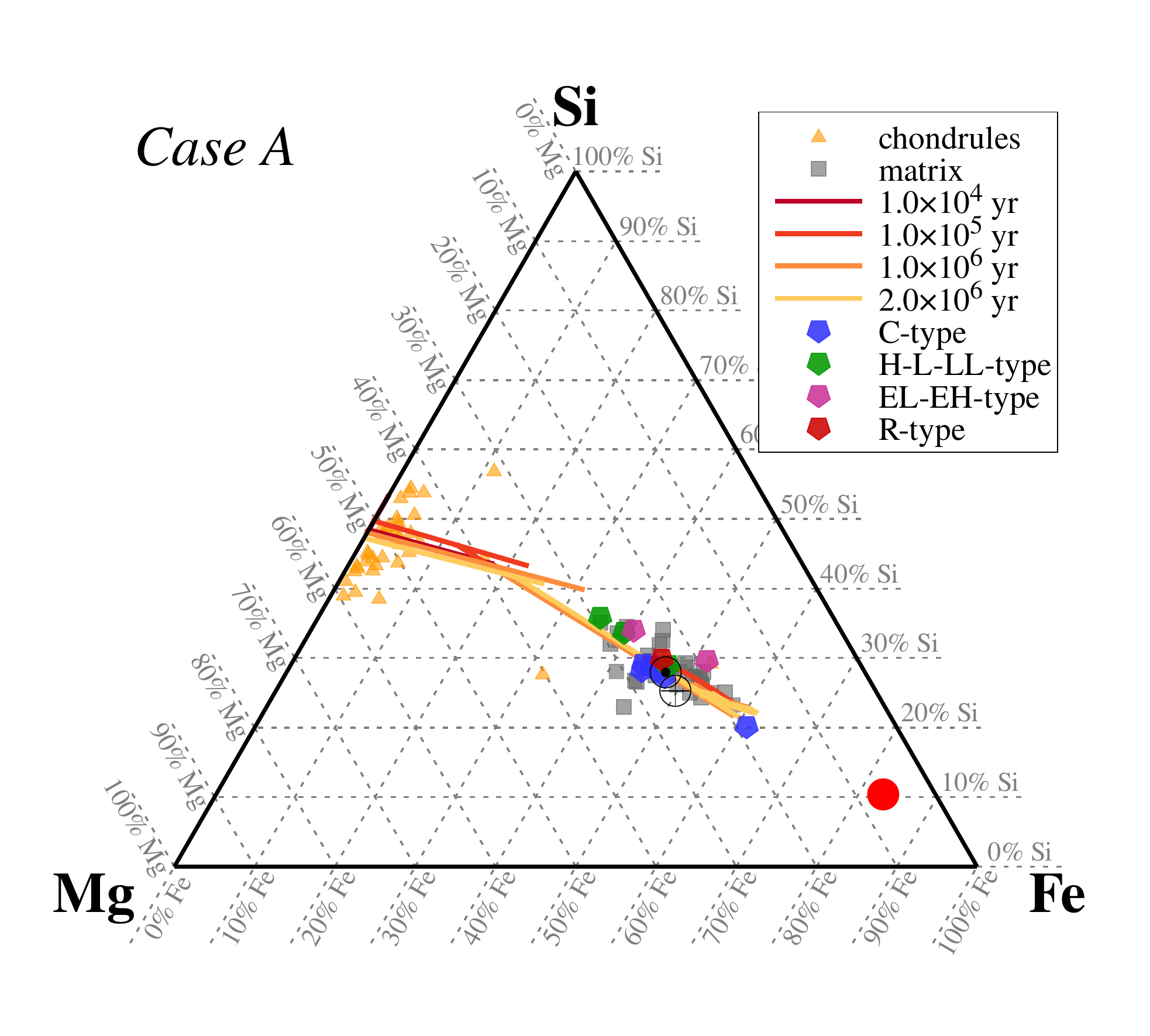}}
		\caption{Ternary diagram (expressed in mass fraction) representing the PSN composition profiles from Fig. \ref{fig:ternaryA} ({\it case A}), with compositions of chondrules, matrixes and mean chondritic types. Compositions of chondrules (yellow triangles) and matrix (grey squares) are taken from \cite{He10}. Mean bulk chondrites compositions (colored pentagons) are taken from \cite{Hu04}.}
		\label{fig:chondrites}
	\end{figure}

	Figure \ref{fig:chondrites} represents the composition of PSN refractory particles in {\it case A} (as in Fig. \ref{fig:ternaryA}) superimposed with the mean bulk compositions of chondrite groups. The figure also shows the composition of a random selection of chondrules and matrix \citep{He10}. The mean bulk composition of chondrites is close to the protosolar value, and its spread seems to follow the profiles derived from our model. The same behaviour can be observed for the matrix. On the other hand, chondrules exhibit a very low amount of bulk Fe, and the average composition seems to be close to the one computed by our model.
	
\begin{figure*}[ht!]
		\resizebox{\hsize}{!}{\includegraphics[angle=0,width=5cm]{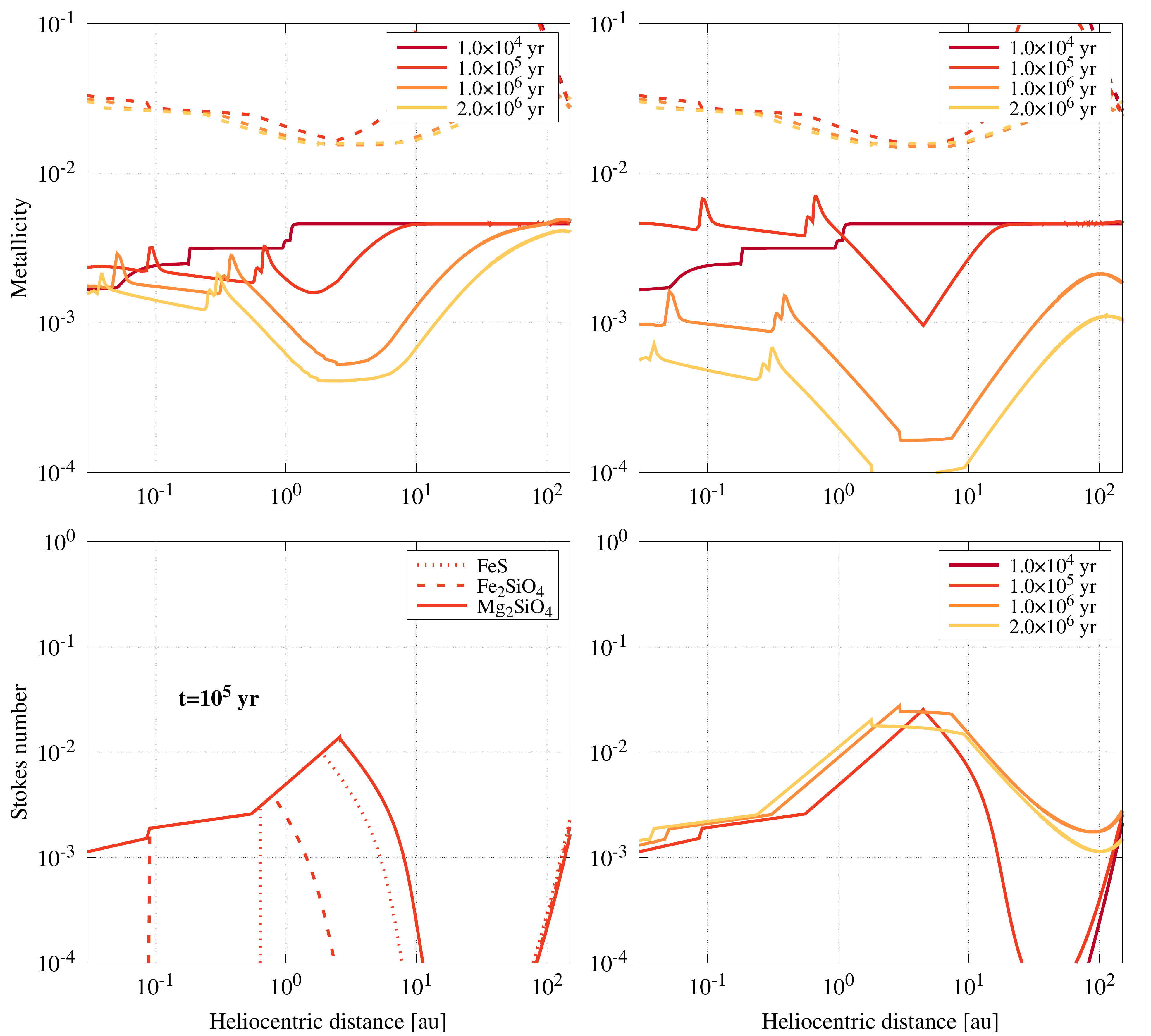}}
		\caption{Local metallicity $Z=\Sigma_{\mathrm{d}}/\Sigma_{\mathrm{g}}$ (top panels) and Stokes number  (bottom panels) computed as a function of time and heliocentric distance. Left and right panels are results for {\it case A} and {\it case B}, respectively (see text). Stokes number is shown at $t=10^6$ yr for a few representative species (left panel) and at different epochs of the PSN evolution (right panel). Dashed lines in top panels show the lowest metallicity $Z_\mathrm{c}$ required to trigger streaming instability (see text).}
		\label{fig:ZSt}
	\end{figure*}

	Top panels of Fig. \ref{fig:ZSt} show the radial profiles of the disk's metallicity (defined as $Z=\Sigma_{\mathrm{d}}/\Sigma_{\mathrm{g}}$) at different epochs of its evolution. As expected, solid matter is concentrated at the position of rocklines. The composition of the PSN around the first cluster of rocklines (iron sulfide, cite and nickel) corresponds to the 30-60 wt\% Fe part of curves in ternary diagrams and Fe wt\% profiles, which matches the S-BO type (barred olivine) spherules compositions. The composition of the PSN around the second cluster of rocklines (fayalite and ferrosilite) corresponds to the 10-30 wt\% Fe part of curves in the ternary diagram and Fe wt\% profiles, matching the S-V type (glass) spherules compositions. In the same manner, we would expect chondrules to be formed in the innermost regions, were sufficient amount of material is present due to continuous drift from the outer disk. At $t=10^5$ yr and 0.67 AU (rockline of iron sulfide), the PSN has 56 wt\% and  58 wt\% of Fe in {\it case A} and {\it case B}, respectively. This increase of the Fe wt\% leads to compositions of the PSN richer in Fe than the protosolar value. 
	
	Finally, bottom panels of Fig. \ref{fig:ZSt} show the Stokes number of dust grains in the disk. Because {\it case A} has many independent species evolving, we only follow forsterite, fayalite and iron sulfide, namely the most, least, and intermediate refractory materials considered in our particles. In {\it case B}, all grains are mixed together, giving a single Stokes number at each heliocentric distance. Using Eqs. (\ref{afrag}), (\ref{adrift}) and (\ref{adf}), we expect i) $\mathrm{St}_\mathrm{frag} \propto 1/T$, ii) $\mathrm{St}_\mathrm{drift}\propto Z/(rT)$ and iii) $\mathrm{St}_\mathrm{df}\propto 1/(\sqrt{r}T)$ (assuming $\left| \frac{\mathrm{d}\ln P}{\mathrm{d}\ln r}\right|^{-1} \propto 1$). In the innermost region, dust size is limited by fragmentation up to $\sim 5$ AU. In the 5-10 AU range, a competition between drift and drift-limited fragmentation sets the dust grains size. Beyond 10 AU dust is in the growth phase.
	
	In {\it case A}, variations in limiting sizes only come from the difference in bulk densities of grains. As a result, different species display the same Stokes number during most of their drift throughout the PSN (see bottom left panel of Fig. \ref{fig:ZSt}). However, because the amount of solid matter decreases below 10 AU, as a result of sublimation and/or radial drift, the Stokes number diminishes as well. In {\it case B}, minor species embedded in large grains are transported more efficiently toward their rocklines. Hence, higher metallicities are found around the rocklines in {\it case B} compared to {\it case A}, at early epochs. In turn, the PSN becomes depleted in solid matter in shorter timescales in {\it case B} compared to {\it case A}.
	
	As indicated in Section \ref{sec:disk-model}, $\alpha$ is a free parameter whose value can change with time and heliocentric distance \citep{Ka15}. For this disk model, an increase of the $\alpha$ value leads to a centrifugal radius $r_\mathrm{c}$ located at higher heliocentric distance, which in turn leads to a larger disk. This also leads to a larger diffusion coefficient $D_i=\nu$ for the trace species. As a consequence, vapors diffuse outward faster and enrich the solid phase more evenly. This results in peaks of abundances that are wider and less prominent than those observed in right panels of Figs. \ref{fig:ternaryA} and \ref{fig:ternaryB}. For the extreme case $\alpha=10^{-2}$, the peaks of abundance are not observable anymore. This shows that the choice on $\alpha$ is critical for both the PSN and trace species evolutions. However, results of simulations with non-uniform $\alpha$ show that this quantity is increasing with heliocentric distance, and takes values of $\sim$ 10$^{-3}$ at 1 AU \cite[see]{Ka15}. Since we are mostly interested in the dynamics of the inner PSN, we adopted this value for our $\alpha$ parameter.
	
	\section{Discussion and conclusion} \label{sec:ccls}
	
	Our model shows that the diversity of the bulk composition of cosmic spherules and chondrules can be explained by their formation close to rocklines, suggesting that solid matter is concentrated in the vicinity of these sublimation/condensation fronts. The slightly lower Fe content of S-type (ordinary) chondrites than for C-type (carbonaceous) observed in Fig. \ref{fig:chondrites} could be due to a partial sublimation mechanism.
	
\begin{figure}[ht!]
		\resizebox{\hsize}{!}{\includegraphics[angle=0,width=5cm]{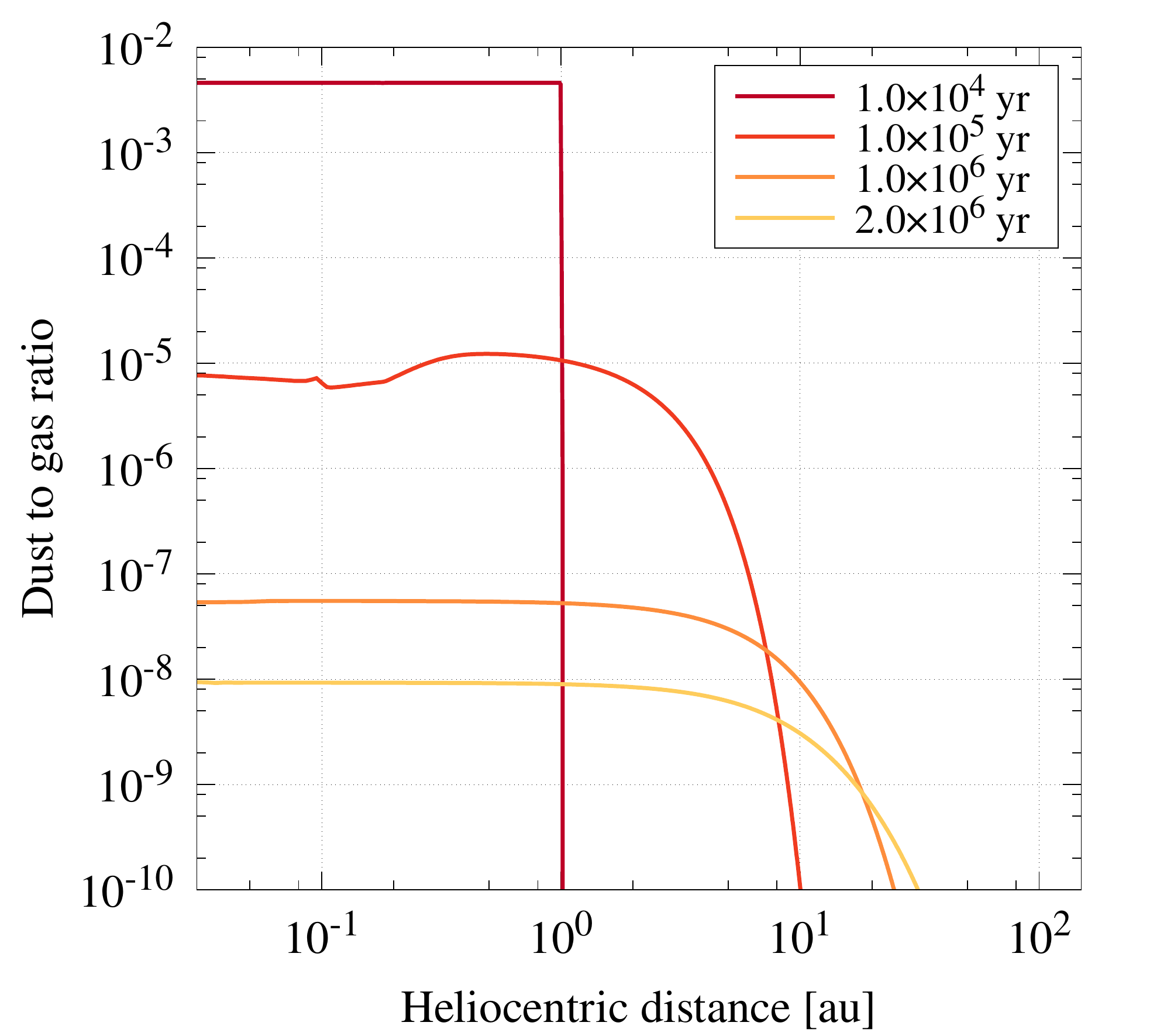}}
		\caption{Time evolution of the radial diffusion of particles in our PSN model, assuming $\alpha=10^{-3}$.}
		\label{fig:dust}
	\end{figure}

	Several transport mechanisms can explain the presence of these processed minerals in the Main Belt, as well as in the outer regions of the PSN. For example, small particles with sizes within the 10$^{-6}$--1~m range and formed in the inner nebula can diffuse radially toward its outer regions \citep{Bo02,De17}. Figure \ref{fig:dust} illustrates the efficiency of diffusion in our model by representing the time evolution of the radial distribution of particles initially formed in the 0--1 AU region of the PSN. In this simulation, a protosolar dust-to-gas ratio is assumed within 1 AU at $t$ = 0 and the effects of rocklines are not considered. The figure shows that particles can diffuse well over 10 AU after 10$^5$ yr of PSN evolution and thus fill the Main Belt region. Photophoresis is another mechanism which can be at play when the inner regions of the PSN become optically thin. In this case, the disk still has a reasonable gas content and allows particles with sizes in the 10$^{-5}$-10$^{-1}$ m range to receive light from the proto-Sun and be pushed outward by the photophoretic force beyond 20 AU in the PSN \citep{Kr07,Mo07,Mo11,Lo16}. These two mechanisms would have taken place before the formation of larger planetesimals in the PSN.
	
	Interestingly, these high concentrations do not allow the triggering of streaming instability. The smallest metallicity $Z_\mathrm{c}$ required to trigger a streaming instability for low Stokes number regimes $\left(\mathrm{St}< 0.1\right)$ is \citep{Ya17}:
	
	\begin{eqnarray}
	\log Z_\mathrm{c} = 0.10 \left(\log \mathrm{St}\right)^2 + 0.20 \log \mathrm{St} - 1.76.  
	\label{zcrit}
	\end{eqnarray}
	
	\noindent Figure \ref{fig:ZSt} shows that the highest disk metallicity computed with our model, i.e. at $t=10^5$ yr and $r$ = 0.1 AU in {\it case B}, is $Z=4.4\times 10^{-3}$, way below $Z_\mathrm{c}=2.7\times 10^{-2}$, suggesting our model does not allow the triggering of a streaming instability. However, the decoupling of dust grains from gas at greater Stokes numbers or the back-reaction of dust onto the gas could slow down the drift of particles in the pile-up regions, thus increasing the local dust-to-gas ratio.
	
	Although the spread in bulk compositions of CS can be well explained by alteration during atmospheric entry \citep{Ru15}, the two scenarios are not mutually exclusive as they happen at very different times. The grains composition computed by our model can be seen as the average composition at each time and heliocentric distance, and the deviation from the mean composition could be the result of full dynamics of grains growth and interaction \cite[same ideas can be found in][]{Ho07}. The effect of rocklines could be among the first processing mechanisms altering the uniform composition of refractory matter in the PSN. Moreover, condensation/sublimation fronts for refractory matter may have other implications. Close to the host star, the pressure at the midplane can be high enough (up to 1 bar) to melt partially or entirely solid grains. This would highly affect the collisional dynamics of grains (or droplets), and allow grains to overcome the meter barrier \citep{Bo14} and form Fe-rich planetesimals that later gave birth to Mercury.
	
	Although our model relies a lot on the number of considered species and the availability of thermodynamic data governing state change, it suggests that rocklines played a major role in the formation of small and large bodies in the innermost regions of the PSN. For example, even if the large amount of iron in Mercury (83 wt\% in the ternary diagram) cannot be explained with this model alone, the increased proportion of Fe in the PSN (62 wt\% at most in the vicinity of rocklines; see Figs. \ref{fig:ternaryA} and \ref{fig:ternaryB}) from the protosolar value (47 wt\%) might have contributed to the accretion of the planet's large core by forming Fe-rich regions. As our model only tracks the evolution of dust grains in the early PSN, it is compatible with current formation mechanisms of terrestrial planets \citep{Ra14,Iz18}. The relevant PSN composition in terms of age and heliocentric distance must then be chosen accordingly to the considered formation scenario.
	
	In its current state, the model fails to reproduce the extreme enrichments in Fe needed to account for the formation of Mercury. However, giant impact simulations performed by  \cite{Ch18} show that the resulting Mercury-like planets display core mass fractions (CMF) in the 0.5--0.7 range (see their Fig. 3), when starting with a protosolar CMF of 0.3. 
		If the initial CMF of Mercury was 10\% higher due to the formation of its building blocks close to rocklines, as suggested by our findings, the post-collision CMF would lie in the 0.6--0.8 range, which is in better agreement with the estimated value of $~0.7$ \citep{St05,Ri07,Ha13}. The combination of multiple scenarios to explain the large CMF of Mercury seems more likely. At greater heliocentric distances the PSN composition becomes again protosolar (mainly for \textit{case B}), which is in agreement with the bulk composition of Earth and Venus (close to protosolar) derived from interior structure models \citep{St05,Rub15,Du17}.
	
	Our study suggests that Mercury-like planets should exist in other planetary systems. More than $\sim$2000 small planets in the 1--3.9 $\Rearth$ range have been discovered so far at close distances to their host stars\footnote{https://exoplanetarchive.ipac.caltech.edu}. Most of the measured densities are poorly determined and the detection of sub-Earth planets remains challenging, implying it remains difficult to quantify the size of the population of Mercury-like planets. Finally, the presence of Mercury-like planets should be ruled by the amount of available matter to form Fe-rich building blocks. In the case of very massive and hot circumstellar disks, rocklines and Mercury-like planets would be located at much higher distances to the host star. In contrast, less massive and colder disks could impede the formation of Fe- rich planets because rocklines would be located too close to their host star.
	
	\acknowledgments
	
	O.M. acknowledges support from CNES. We thank the anonymous referee for useful comments that helped improving the clarity of our paper.

	

\end{document}